\documentclass[twocolumn,aps,prb,showpacs]{revtex4-1}

\usepackage{graphicx}
\usepackage{hyperref}
\usepackage{epstopdf}
\usepackage{amsmath}
\usepackage{amssymb}
\usepackage{upgreek}
\usepackage[tight,TABTOPCAP]{subfigure}
\usepackage{float}
\usepackage{graphicx}
\usepackage{natbib}
\usepackage{color}
\usepackage[usenames,dvipsnames]{xcolor}

\makeatletter

\newcommand{\vect}[1]{{\boldsymbol{#1}}}

\begin{document}

\title{Equation of State of the Fermionic 2D Hubbard Model}

\author{J. P. F. LeBlanc$^{1}$}
\email{jpfleblanc@gmail.com}
\author{Emanuel Gull$^{2}$}

\affiliation{$^1$Max-Planck-Institute for the Physics of Complex Systems, 01187 Dresden, Germany} 
\affiliation{$^2$Department of Physics, University of Michigan, Ann Arbor, MI 48109 USA}


\date{\today}
\begin{abstract}
We present results for the equation of state of the two-dimensional Hubbard model on an isotropic square lattice as obtained from a controlled and numerically exact large-cluster dynamical mean field simulation.  Our results are obtained for large but finite systems and are extrapolated to infinite system size using a known finite size scaling relation, and are supplemented by reliable error bars accounting for all sources of errors.
We establish the importance of examining the decay of spatial spin correlations to determine if a sufficiently large cluster has been used and with this in mind we present the energy, entropy, double occupancy and nearest-neighbour spin correlations extrapolated to the thermodynamic limit. We discuss the implications of these calculations on pseudogap physics of the 2D-Hubbard model away from half-filling, where we find a strong behavioural shift in energy below a temperature $T^*$ which becomes more pronounced for larger clusters.  Finally, we provide reference calculations and tables for the equation of state for values of doping away from half-filling which are of interest to cold atom experiments.
\end{abstract}

\maketitle
\section{Introduction}
The single-orbital Hubbard model in two dimensions is one of the simplest models of correlated electron physics: it describes electrons on a lattice moving with a hopping strength $t$ between nearest neighbor sites and interacting with an interaction strength $U$ if two electrons are on the same site.
The model is known to have a Fermi liquid phase at weak interaction strength and low doping, an insulating phase with a large gap at half-filling and large interaction strength, and a d-wave superconducting phase in at least some part of parameter space.\cite{zanchi:1996,maier:2005b} A `pseudogapped' phase also exists near half-filling in which the electronic spectrum is strongly suppressed around the antinode but not along the nodal direction.  The model has been realized in cold fermionic gas systems and some of these phases, in particular the Mott-insulating state, have been observed experimentally.\cite{jordens:2008,schneider:2008}  
It is also believed that the essence of the physics responsible for superconductivity in the high transition temperature ($T_c$) cuprate superconductors stems from the strong correlations described in the 2D Hubbard model for intermediate values of $U$.\cite{anderson:1987, lee:2006, basov:2005}
  This is particularly evident in the underdoped region of the hole-doped cuprate phase diagram where there exists a pseudogap phenomenon thought to emerge from strong correlation physics as the system is doped away from the Mott-insulator at half-filling.\cite{timusk:1999}  Pseudogap-like spectra have been observed in a wide range of approximate analytical\cite{Kampf90,Castellani95,Wen96,Abanov01,lee:2006} and numerical calculations.\cite{huscroft:2001,macridin:2006,gull:2008,werner:2009,gull:2009,gull:2010,Lin10,sakai:2012}
However, the cold gas experiments, which attempt to replicate the physics of the model with ultra cold fermions, are so far unable to reach temperatures low enough to show subtle correlation physics.\cite{jordens:2010}
 
Standard analytical techniques applied to correlated electron systems have not been able to provide reliable and unbiased results for the equation of state, phases, or phase boundaries in the correlated intermediate coupling regime relevant to the interesting cuprate physics.  These techniques can be successful, however, in limits where the Hamiltonian can be expanded in orders of some small parameter.  One example is the high temperature series expansion (HTSE) which is based on an expansion of the Hamiltonian in powers of the inverse temperature $\beta$.  Because of this small parameter limitation to analytic work, insight into the physics of the Hubbard model must therefore come from numerical simulations\cite{scalapino:2007} that are able to access the correlated regime in a controlled way.\cite{dagotto:1994,rigol:2006,rubtsov:2008,rohringer:2012,kozik:2010, hirsch:1985, hirsch:1989} Several candidates which are either exact or very accurate in some region of the phase diagram exist.  
One technique that provides results directly in the thermodynamic limit is the numerical linked-cluster expansion (NLCE).  For this model it is accurate\cite{khatami:2011} at high $T$ and large $U$ but the results diverge at low temperature and weaker $U$.  Data beyond this divergence can only be obtained with the use of approximate numerical resummation techniques\cite{tang:2012}  which lack a small parameter and are therefore uncontrolled.  Another technique, variational Monte Carlo, is based on approximating the true ground-state wave function at zero temperature by a variationally optimized trial wavefunction.\cite{Gros88,Yokoyama88,Becca00,Paramekanti01,Paramekanti04,Yokoyama04,Ogata12a,Ogata12b}  Other Monte Carlo methods, such as Lattice (`determinantal') quantum Monte Carlo (DQMC)  are numerically exact when combined with both a lattice finite size and a Trotter extrapolation. However, they encounter a severe sign problem away from half-filling. Gaussian\cite{Corney04}, diagrammatic\cite{kozik:2010} and bold-line\cite{Prokofev08} Monte Carlo methods have been proposed and are currently under investigation.

Away from weak or strong coupling and away from high symmetry points (e.g. half-filling), the equation of state of the Hubbard model is only known at high temperature.\cite{khatami:2011}
In this work we change this situation by providing the numerically exact equation of state, with error bars, for the two-dimensional Hubbard model for interaction strengths ranging from weakly to strongly coupled, with an emphasis on doping near to half filled. Our goals are threefold:
First, to provide a numerically exact equation of state in regions that were previously inaccessible. Second, to provide reference data for use in experimental systems trying to replicate Hubbard model physics, {\it e.g.} cold atomic gas systems, and third, to provide reliable comparison and benchmark data to which new numerical and analytical methods can be compared and for which their reliability can be tested.

To accomplish these goals we employ the dynamical cluster approximation (DCA), one of several cluster extensions\cite{Hettler98,Hettler00,Lichtenstein00,Kotliar01,maier:2005} to the dynamical mean-field theory (DMFT).\cite{Metzner89,Georges92b,georges:1996} 
DCA is a controlled technique based on a finite size cluster embedded in a bath, which has the number of cluster sites as a small parameter. DCA on any finite cluster also provides the full frequency dependence of the Green's function and self energy, but approximates its momentum dependence.
Using the convergence of the DCA to the thermodynamic limit (TL) as a function of its small parameter we obtain converged lattice self energies and single particle Green's functions for the 2D Hubbard model and  compute the equation of state over a range from high temperature, $T\approx 10 t$, down to intermediate temperature, $T \approx 0.3 t$.  We explore $U$=4, 8, 12 for weak, intermediate, and strong coupling as well as a range of doping away from half-filling from $n=0.85$ to 1.0. Where controlled high-temperature results from NLCE are available, we compare to these. We also show lower temperature extrapolated NLCE results at select places.

We present the essential theory and outline the computational technique used in Sec.~\ref{sec:theory}.  Sec.~\ref{sec:results} will contain our main results and discussion while Sec.~\ref{sec:conclusions} will conclude.  A database of numerical results for the equation of state of the Hubbard model along with a detailed description of these results is included in the supplementary material.\cite{SM}

\section{Theory}\label{sec:theory}
The Hubbard model Hamiltonian is given by
\begin{equation}\label{eqn:H}
H=-\sum\limits_{ \langle i,j\rangle \sigma}t \left( c_{i\sigma}^\dagger c_{j\sigma} + h.c.\right) +U\sum\limits_i n_{i\uparrow}n_{i\downarrow},
\end{equation}
where $c^\dagger_{i\sigma}$ and $c_{i\sigma}$ create and annihilate (respectively) an electron with spin $\sigma=\uparrow$,$\downarrow$ on site $i$,  $n_{i\sigma}=c^\dagger_{i\sigma}c_{i\sigma}$ is the number operator, and  $\langle i,j\rangle$ denotes a summation over nearest neighbour pairs with nearest neighbour hopping energy, $t$, which sets the scale of all energies presented in this work.

\begin{figure}
  \begin{center}
  \includegraphics[width=\linewidth]{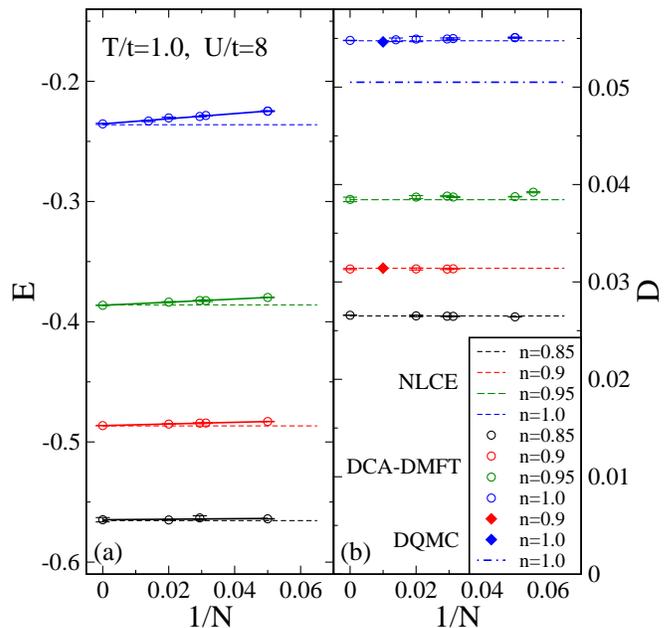}
  \end{center}
  \caption{\label{fig:extrap}(Color online)(a) $E=E_K+E_V$ in units of $t$, plotted as a function of the inverse cluster size, $1/N$, for $T/t=1.0$ and $U/t=8$ at densities, $n$, near half-filling.  Horizontal dashed lines are results from NLCE data.\cite{ehsan:priv}  (b) Double occupancy at half-filling for $T/t=1.0$ and $U/t=8$.  Horizontal dashed curves are NLCE data\cite{ehsan:priv} and colored diamonds are values from determinantal quantum Monte Carlo (DQMC) results at and away from half-filling\cite{gorelik:2011, scalettar:priv} and the dashed-dotted line is the value extracted from DQMC results of Ref.~\onlinecite{paiva:2010} [Fig.2(c)] at half-filling.  }
\end{figure}  
\begin{figure}
  \begin{center}
  \includegraphics[width=\linewidth]{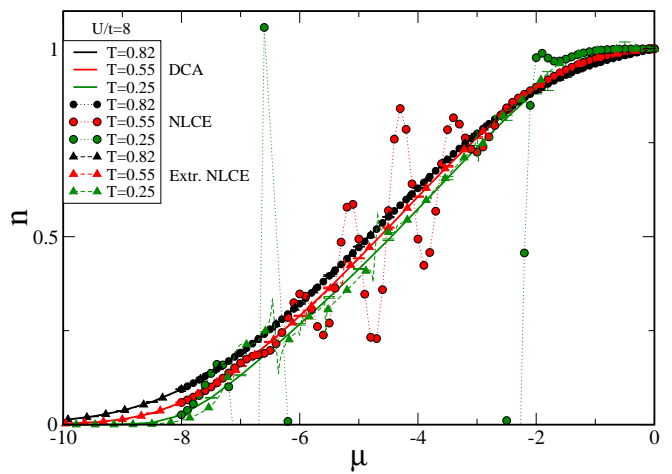}
  \end{center}
  \caption{\label{fig:nmu}(Color online) The electron density per lattice site, $n$, as a function of chemical potential, $\mu$, for the intermediate coupling case of $U/t=8$ for $T/t=$0.82, 0.55, 0.25.  We choose $\mu$ relative to $U/2$ so that half-filling corresponds to $\mu=0$. Results from NLCE and the extrapolated resummations are included for comparison.\cite{khatami:2011, ehsan:priv}   }
\end{figure}

We solve the model in the `dynamical cluster approximation' (DCA). Within DMFT,\cite{Metzner89,Georges92b,georges:1996} the self energy is approximated as a local, momentum-independent, quantity.  This allows one to map the problem to the solution of an auxiliary Anderson impurity model (AIM) of a local impurity in a self-consistently adjusted mean field instead of the numerically intractable infinite lattice model. Cluster extensions are then used to systematically reintroduce some momentum- and frequency dependence of the self energy.\cite{Fuhrmann07} Within DCA,
\begin{equation}\label{eqn:dca}
\Sigma(\vect{k}, \omega)=\sum\limits_{K=1}^N \phi_{K}(\vect{k})\Sigma_{K}(\omega)
\end{equation}
where $\vect{k}$ is the momentum, $\omega$ is frequency and $K$ is a label for each of the N patches in a cluster.  $\phi_K(\vect{k})$ is taken to have value $1$ for a momentum $\vect{k}$ which lies in momentum patch $K$, and zero for any $\vect{k}$ outside of this patch.  Hence, the DCA approximation to the self energy is a piecewise constant function, though other forms have been attempted.\cite{ferrero:2009,staar:2013} As $N\to \infty$, the DCA momentum-space patchwork becomes a continuum of states providing exact momentum and frequency dependencies to the self energy and Green's function.   In the following we will present results for the energy, entropy, nearest and further neighbour spin correlations as well as specific heat obtained by DCA on a finite cluster and then extrapolated to the TL. The kinetic and potential energies can be obtained from \cite{georges:1996, gull:2012}
\begin{align}\label{eqn:nrg}
E_K &=\sum\limits_{\boldsymbol{k}\sigma}(\epsilon_{\boldsymbol{k}}-\mu)\langle c_{\boldsymbol{k}\sigma}^\dagger c_{\boldsymbol{k}\sigma} \rangle\nonumber \\&=  2T\sum\limits_{\boldsymbol{k},n}(\epsilon_{\boldsymbol{k}}-\mu){\rm Tr}[G(\boldsymbol{k},i\omega_n)],\\
E_V &= U \sum\limits_{i}\langle n_{i\uparrow}n_{i\downarrow} \rangle \nonumber \\&= T \sum\limits_{\boldsymbol{k},n} {\rm Tr} [\Sigma(\boldsymbol{k},i\omega_n)G(\boldsymbol{k},i\omega_n)],\label{eqn:vnrg}
\end{align}
which are summed over momentum, $\vect{k}$, and fermionic Matsubara frequencies, $i\omega_n= (2n+1)\pi/\beta$, and where $\epsilon_{\vect{k}}=-2t(\cos k_x a + \cos k_y a)$ is the tight binding dispersion for the simple square lattice with lattice constant $a$.

To solve the impurity problem we use the continuous-time auxiliary field algorithm,\cite{gull:2008} a continuous-time method\cite{gull:2011, rubtsov:2005} with sub-matrix updates\cite{gull:2011:smu}  which allows for numerically exact solutions of large clusters.\cite{fuchs:2011} 
In 2D, the convergence of an integral over a discretized periodic function as the discretization becomes finer goes like $(N^{1/2})^{-2}$.  We therefore expect a linear convergence as a function of $1/N$ of local quantities to the infinite cluster size, $1/N\to0$. In this paper we extrapolate using clusters of $N= 20, 32, 34$ and $50$ to determine all TL quantities unless otherwise noted.
We present extrapolated data except where otherwise noted and provide an example of convergence in Fig.~\ref{fig:extrap} as well as include the data for finite cluster sizes in the attached supplementary material.\cite{SM}  Away from half-filling a sign problem occurs.\cite{Troyer05}  This is most dominant in CT-AUX in a range of $n=0.8\to1.0$.  For these densities, once $T/t\approx 0.6$ the sign problem begins and further reduction in temperature is exponentially more computationally intensive.

The entropy for a given temperature, $T$, and doping $n$ is obtained from the total energy through
\begin{equation}
S(T , n)=S(T_u, n) + \frac{ E(T , n)}{T} - \int \limits_T^{T_u} \frac{E(T^\prime , n)}{{T^\prime}^2}dT^\prime
\end{equation}
where $S(T_u, n)$ is the high temperature, $T_u$, limit which acts to offset the entropy such that $S(T=0)=0$.  Here we take $T_u/t=10$ and $S(T_u,n)$ from NLCE data.\cite{ehsan:priv}  The specific heat can be obtained from the energy without dependence on this constant offset through the derivative $C(T,n)=\frac{\partial E(T)}{\partial T}$.

The DCA construction of Eq.(\ref{eqn:dca}) provides momentum space variation in the self energy and Green's function.   By Fourier transforming to real space we can extract information on a length scale smaller than the cluster size in addition to thermodynamic properties. This is done at the cluster level during the DMFT loop in the QMC impurity solver.  We measure the average occupancy on each lattice site $\langle  n_{i\sigma} \rangle$, as well as the average correlated occupancies $\langle n_{i\sigma} n_{j\sigma^\prime} \rangle$.  From these we obtain quantities of interest such as the density per lattice site, $n=\langle n_{i\uparrow} +n_{i\downarrow} \rangle$, the double occupancy, $D=\langle n_{i\uparrow}n_{i\downarrow} \rangle$, and the spin correlations  $\langle S_i^z S_j^z\rangle=\langle (n_{i\uparrow}-n_{i\downarrow})(n_{j\uparrow}-n_{j\downarrow}) \rangle$.\cite{paiva:2001}  We present such quantities extrapolated to the TL.

We also provide estimates of the uncertainty of our calculations. For any finite cluster size, the only error is the stochastic Monte Carlo error of our quantum Monte Carlo impurity solver, which decreases as the inverse of the square root of our computational time. For all observables we estimate our uncertainties from the statistical variation between independent Monte-Carlo iterations in a converged DMFT loop. For functions of observables ({\it e.g.} the energy), we apply a jack-knife procedure.
For quantities which are extrapolated to the thermodynamic limit, we show the error obtained by a linear regression analysis of the extrapolation to infinite system size. In this case the errors in extrapolated values represent only the scatter of the various cluster sizes.  This is useful as it gives a measure of the quality of the extrapolation, and is reasonable where the statistical fluctuations are much smaller than the finite size scatter.  This is the case for most of our calculations, so that we expect that our error is well represented by the linear regression error.


\section{Results and Discussion}\label{sec:results}
In order to establish the validity of the finite size extrapolation we first show for a single temperature the calculation of the total energy, $E=E_K+E_V$ where the kinetic ($E_K$) and potential ($E_V$) energies are given by Eqns.~(\ref{eqn:nrg}) and (\ref{eqn:vnrg}). 
The total energy is plotted in Fig.~\ref{fig:extrap} as a function of inverse cluster size, $1/N$,  at a fixed $T/t=1.0$ and $U/t=8$.   
Horizontal lines in Fig.~\ref{fig:extrap} represent NLCE reference data of Ref.~\onlinecite{khatami:2011} which, at these temperatures and fillings,  agree precisely with the extrapolated DCA values.
Fig.~\ref{fig:extrap}(b) shows similar extrapolation in $1/N$ for the double occupancy, $D=\langle n_{i\uparrow}n_{i\downarrow}\rangle$, again for fixed $T/t=1.0$ and $U/t=8$.  
Dashed horizontal lines again represent the NLCE data at $U/t=8$, and agree with the present DCA extrapolation.  
Our DCA data disagrees with the lattice Monte Carlo calculations of Ref.~\onlinecite{paiva:2010}, shown as the horizontal dashed-dotted line, on the 10\% level.  However, it is in perfect agreement with more recent DQMC data.\cite{scalettar:priv,gorelik:2011} We believe that the discrepancy with Ref.~\onlinecite{paiva:2010} is caused by a finite $\Delta \tau$ Trotter error in the Monte Carlo process in imaginary time which occurs in that work but which could, in principle, be controlled.\cite{Paiva11} 

In Fig.~\ref{fig:nmu} we show the density $n$ per lattice site for a fixed $U/t=8$.   The solid lines are the DCA extrapolations in $1/N$ of cluster sizes $N=20,32, 34, 50$ to the thermodynamic limit.  
At high temperature, $T/t=0.82$, we see that the DCA results agree within error with the extrapolated results of the numerical linked-cluster expansion calculations\cite{khatami:2011, ehsan:priv} shown as triangular black points, validating the extrapolation that has been used at high $T$.
At high temperature the NLCE agrees with our DCA results.  As temperature is reduced, there is an intermediate regime where the NLCE begins to diverge but through numerical resummation can be extrapolated to the correct value which agrees with our numerically exact DCA calculations.  At low temperature this extrapolation of NLCE data fails while DCA remains accurate.
Shown for $T/t=0.55$ and more clearly at $T/t=0.25$, there is a large range of $n$ which is not consistently accessible by NLCE.  This is in contrast to the DCA technique which can consistently access a broad chemical potential range at these temperature. 

For large $U$ as temperature is reduced one can see the formation of an incompressible region near half-filling which occurs in a range of $\mu$ around $\mu=-1 \to 0$ in the $T/t=0.25$ case of $U=8$ shown in Fig.~\ref{fig:nmu}.\cite{paiva:2010, khatami:2011}  
This behaviour characterizes the Mott state at half-filling which has previously been examined in DQMC\cite{paiva:2010} and in cluster DMFT on smaller clusters.\cite{gull:2009,gull:2008b,Parcollet04}
 In Ref.~\onlinecite{paiva:2010} the authors traced the range of $\mu$ over which the density was incompressible.  This range was then interpreted as a measure of the size of a gap in the density of states.  
In DCA we find a momentum dependent pseudogap  where this incompressibility occurs only in the antinodal regions of the Brillouin zone in addition to the interesting Mott physics at low temperature.\cite{gull:2010}  The onset of this momentum selective incompressibility with temperature is also signified by a peak in the spin susceptibility at $T^*$, which has previously been interpreted as the pseudogap onset\cite{paiva:2010} and shown to coincide with the formation of a pseudogap in the density of states.\cite{macridin:2006} 

\begin{figure}
  \begin{center}
  \includegraphics[width=0.95\linewidth]{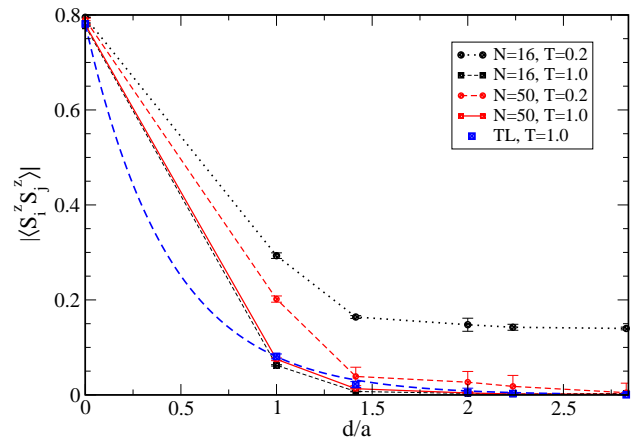}
  \end{center}
  \caption{\label{fig:sisj}(Color online) The magnitude of the spin-spin correlation function $\left|\langle S_i^z S_j^z\rangle\right|$ at half-filling as a function of distance, $d/a$, for $T/t=0.2$ (circles) and 1.0 (squares), for a 16-site (black online) and 50-site (red online) cluster.  For $T=1.0$ the extrapolation to the TL is shown, with exponential fit $|\langle S_i^z S_j^z\rangle| = A e^{-d/\xi}$ with $\xi\approx0.43$ and $A=\langle {S_i^z}^2\rangle\approx 0.78$. }
\end{figure} 
%

 One expects that as temperature is lowered the  length scale of correlations in the system should grow.  To ensure that our clusters have sufficient size to account for this increasing correlation length, we increase the cluster size until we see convergence in a quantity of interest.  
\begin{figure}
  \begin{center}
  \includegraphics[width=0.95\linewidth]{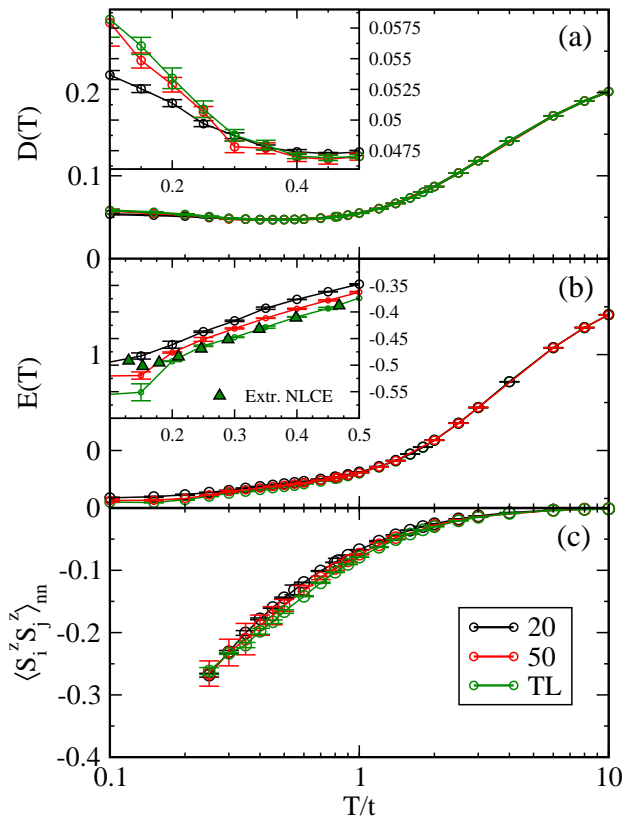}
  \end{center}
  \caption{\label{fig:D}(Color online) Double occupancy, $D(T)$, energy, $E(T)$, and nearest neighbour spin correlations at half-filling are shown in frames (a), (b) and (c) respectively as functions of $T/t$ at half-filling for $U/t=8$.  Results are shown for 20 (black online) and 50 (red online) site cases as well as extrapolations to the TL (green online) as described in the text.  Inset of (a) and (b) are enlargements of the low temperature regions of their respective figures.  The inset of (b) includes extrapolated NLCE data.\cite{ehsan:priv, khatami:2011} }
\end{figure}  
The spin-correlation function, $\langle S_{i}^z S_j^z \rangle$, as a function of distance, $d=|\boldsymbol{x}_i-\boldsymbol{x}_j|$, is such a quantity.  Since the system is antiferromagnetic we remove the alternating sign and instead plot the magnitude for each neighbour distance, $\left| \langle S_{i}^z S_j^z \rangle \right|$ in Fig.~\ref{fig:sisj}.  
As can be seen at high temperature, $T/t=1.0$, there is little variation between a 16-site and 50-site calculation of spin correlations, shown as black and red squares.  
One can see that all relevant correlations are accounted for, as the amplitude of $\left| \langle S_{i}^z S_j^z \rangle \right|$ decays to zero within the linear cluster size of the small cluster.  Also shown is the extrapolation of $\left| \langle S_{i}^z S_j^z \rangle \right|$ to the thermodynamic limit for $T/t=1.0$, which can be reasonably fit by an exponential decay $\left| \langle S_{i}^z S_j^z \rangle \right| = \langle  {S_{i}^z}^2\rangle e^{-d/\xi}$ as expected from analytic work on the 2D Hubbard model.\cite{auerbach} The spin-spin correlation length fitting results in $\xi\approx 0.43$ which is smaller than half the linear cluster size for both the 16 and 50-site cases.  Thus there is no new physics in the 50-site case which is not in the 16-site case at $T/t=1.0$.  Any difference in results for thermodynamic properties for increasing cluster size must be perfectly accounted for by the $1/N$ DCA scaling.
The utility of this analysis becomes apparent at low temperatures, illustrated here at $T/t=0.2$, again for the $N$=16 and $N$=50 cases.  While the $d=0$ on-site correlations remain unchanged, the non-local correlations differ drastically between the two cluster sizes.  
  Regardless of the physical or computation source of this cluster size discrepancy the examination of the spin-spin correlations gives an excellent metric to determine if sufficiently large clusters have been included and allows us to overcome this issue at low temperature by extrapolating only with clusters large enough to include all relevant correlations.  For the data presented in this work, this will manifest as a natural minimum accessible temperature based on our maximum cluster size of 50 sites.  This minima can be overcome by extrapolating with larger cluster sizes, but the precise clusters required become a detail of the observed spin correlations as in Fig.~\ref{fig:sisj}.
   More importantly, from this, one can see to what level spin-spin correlations are maintained in various cluster sizes.  
  
In the infinite $U$ limit a system should contain no double occupancy at half-filling but for any finite $U$ this is not the case.
In Fig.~\ref{fig:D}(a) we show the double occupancy obtained from clusters of size $N=$20, 50 and in the extrapolation to the TL.  At very low temperature we see in the case of $N$=20 that the expected reduction in double occupancy for reduced temperature begins to reverse below $T\approx0.5$.\cite{paiva:2010}  We also note (see inset) that as we push towards $N$=50 and the TL that the double occupancy at low temperature increases further.
The rise in double occupancy, which is related to the potential energy, coincides with a continually decreasing total energy shown in the inset of Fig.~\ref{fig:D}(b).  This indicates a reduction in kinetic energy which allows us to understand the rise in double occupancy as a physical consequence of the electrons becoming localized.
This same effect has been phrased previously as a consequence of a low temperature increase in the local spin moment, $\langle {S_i^z}^2\rangle$ for reduced temperatures caused by a rise in double occupancy.\cite{paiva:2001}
Examining higher temperature there is a behavioural shift in the double occupancy.  This occurs in the range $T/t=1.0 \to 2.0$ where the double occupancy changes from the roughly constant value of $D=0.05$ to having a continued increase with temperature.
With this in mind we can examine the low temperature behaviour of the energy in Fig.~\ref{fig:D}(b).  At high temperatures we see a rise in energy which mimics the rise in $D$ above $T/t=2.0$.  At low temperatures we see the need for large cluster sizes.   For the smaller cluster of $N=20$ the energy is nearly smooth to temperatures as low as $0.1t$.  We see however a shift which occurs only at low temperatures in the large clusters.  This shift occurs at a temperature which reasonably agrees with the previously identified pseudogap temperature scale, $T^*\approx 0.3t$,\cite{paiva:2010} the temperature below which a reduction of the density of states is observed to occur in the antinodal direction but not in the nodal direction.\cite{macridin:2006, paiva:2001} While this feature is present in the 20-site case it is only extremely weak and, with the exception of Ref.~\onlinecite{mikelsons:2009}, has been mostly unmentioned in previous works which considered only smaller clusters.

We note the agreement of our results with extrapolated NLCE data\cite{ehsan:priv} shown for intermediate temperatures in the inset of Fig.~\ref{fig:D}.
We also examine the spin correlations, $\langle S_i^z S_j^z \rangle_{nn} $ over the set of nearest-neighbours (nn) with variation in temperature plotted in Fig.~\ref{fig:D}(c).   
However, in this case we have omitted the data points at the lowest temperatures since there the uncertainties in this quantity become too large for a reliable extrapolation.
 
\begin{figure}
  \begin{center}
  \includegraphics[width=0.95\linewidth]{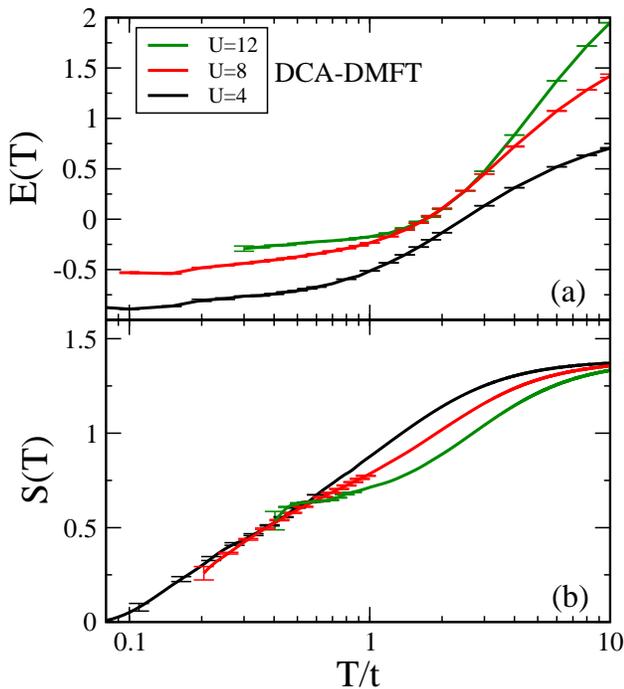}
  \end{center}
  \caption{\label{fig:esc}(Color online) Energy, $E(T)$, and entropy, $S(T)$, as functions of $T/t$ at half-filling extrapolated from DCA data to the TL for $U/t$=4, 8 and 12.}
\end{figure}  

In Fig.~\ref{fig:esc} we present results of the energies from DCA extrapolated to the TL for varied interaction strength at half-filling.  As is apparent in other works,\cite{Parcollet04,gull:2008b,gull:2009,khatami:2011} the $U=4$ case does not show an incompressible phase at these temperatures. 
It is expected that the incompressible regime will have some impact on the intermediate and strongly coupled energies.  While the effect is subtle in the energy, the cumulative effect on the entropy, shown in Fig.~\ref{fig:esc}(b), results in a decrease in $S(T)$ below $T^*$.  On physical grounds this represents the loss of available thermal configurations at finite temperature as the electronic density of states becomes gapped and enters a partially gapped pseudogap state.  This momentum-selective Mott transition is the same physics which explains the partially incompressible region of densities near half-filling at, for example, $T=0.25$ in Fig.~\ref{fig:nmu} and may may have consequences for the interplay between superconductivity and the pseudogap.\cite{gull:2012,gull:2009, werner:2009, gull:2010} Though such a depression exists in the strong coupling case for N=50 we cannot accurately extrapolate to the TL below this temperature with our current range of cluster sizes and limit our present work to $T>0.3t$ at the value of $U/t=12$.



\begin{figure}
  \begin{center}
  \includegraphics[width=\linewidth]{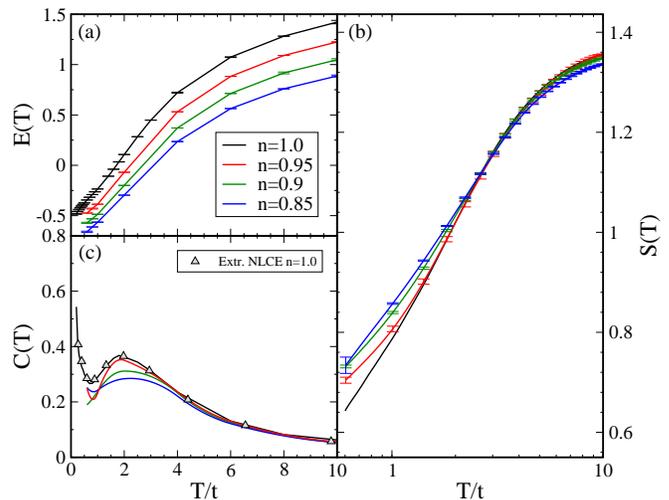}
  \end{center}
  \caption{\label{fig:u8filled}(Color online) Energy, $E(T)$, entropy, $S(T)$, and specific heat capacity, $C(T)$, as functions of $T/t$ extrapolated to the TL for $U/t=8$ for filling values of $n=0.85$, 0.9, 0.95, and 1.0 (half-filled).  The extrapolated NLCE data in (c) can be found in Ref.~\onlinecite{khatami:2012}.}
\end{figure}

In Fig.~\ref{fig:u8filled}(a) and \ref{fig:u8filled}(b) we present for $U/t=8$ the energy and entropy respectively for doping values near to but away from half-filling.
In addition, the energies also provide direct access to the electronic specific heat shown in Fig.~\ref{fig:u8filled}(c). 
Our $C(T)$ data are obtained by taking finite differences in the spline interpolation of neighbouring energy values and therefore amplifies the numerical noise of Fig.~\ref{fig:u8filled}(a).  For $C(T)$ we omit error bars as the value and uncertainty are somewhat dependent upon the method of interpolation and differentiation.  Despite this, our results agree with the extrapolated NLCE data at half filling.\cite{khatami:2012} We have also extended the present work to include three dopings, of $n=0.85$, 0.90 and 0.95, away from half filling in the region most difficult for DCA calculations due to the occurrence of a sign problem.  Here finite size issues in DCA result in deviations from standard DMFT results.  Though not explored here, this present work shows that coarsely gridded and interpolated DCA data can be used to obtain precise specific heat data at dopings far away from half-filling, where other techniques cannot converge at low temperatures.
 Other Monte-Carlo works \cite{duffy:1997,paiva:2010, fye:1987} obtained on finite systems have identified the two main features of the specific heat, namely the low temperature spin and high temperature charge peaks.    Here we present accurate results of the high temperature charge peak (near $T/t=2.0$) in the TL.  At low temperatures we simply remark that the impact of the shift in energy, which is only apparent for large clusters, acts to create the spin peak in $C(T)$.\cite{khatami:2012}  For cold-atom experiments both of these peaks in $C(T)$ will act as a strong barriers to further cooling of an atomic gas system.  

The results presented in Figs.~\ref{fig:extrap} to \ref{fig:u8filled} include only a small part of the numerical results which we make available in this paper.  For the sake of brevity we organize these additional results in the supplementary material which contains a detailed explanation of the data sets.  In addition to the $U/t=8$ data we have presented here, we also include in the supplement the extrapolations to the thermodynamic limit for $U/t=4$ and 12 both at and away from half-filling.  We expect these results to be a useful reference for comparison with other techniques in parts of phase space (in particular at low $T$, away from half-filling) where no previous controlled Monte-Carlo results exist.

\section{Summary and Conclusions} \label{sec:conclusions} 
  
We have calculated the full thermodynamics of the 2D-Hubbard model by extrapolating DCA results on large clusters to the thermodynamic limit.  Our results are numerically exact and, at high temperature, are validated against numerical linked-cluster expansion results.  We have extended our parameter range substantially beyond what was previously shown.  We provide results in the thermodynamic limit, for lower temperatures as well as for a wide range of filling values.  We assert  that our results are numerically  exact within the errors we provide, verified by explicitly examining the range of spin correlations in real space.  From this we can observe that our choice of cluster sizes has included all correlations.  

We note the occurrence of low temperature features in energy and entropy which seem to correlate with the onset of pseudogap physics at $T^*$ which are not captured directly in thermodynamic quantities for small clusters. Finally, we present exact results for nearest-neighbour spin correlations.  Since $\langle  S_i^z S_j^z\rangle_{nn}$ is measurable in cold-atom experiments, it may be used for thermometry.\cite{greif:2011,kollath:2006}  Accurate values and reliable error bars are essential for this purpose.  We have shown that DCA is an ideal technique for establishing the temperature dependence of these correlations, and have provided tables in the supplement which contain reference data needed for alternate techniques.


\begin{acknowledgments}
We thank Ehsan Khatami for useful discussions and the extrapolated NLCE reference data we present in this paper, and Richard Scalettar for investigating the discrepancy with Ref.~\onlinecite{paiva:2010} and providing us with updated DQMC data.\cite{scalettar:priv}  Our continuous-time QMC codes are based on the ALPS libraries.\cite{gull:2011:alps, bauer:2011}
\end{acknowledgments}


\bibliographystyle{apsrev4-1}
\bibliography{bib}

\end{document}